\documentclass[11pt,a4paper]{article}
\pdfoutput=1

\usepackage[T1]{fontenc}
\usepackage[utf8]{inputenc}
\usepackage{jcappub}

\graphicspath{{./}}

\title{Compact stars in a large-tension braneworld: mildly negative Weyl coupling consistent with NICER and gravitational-wave data}

\author[a]{Samuel Isidoro dos Santos J\'unior}
\author[a]{Rafael Camargo Rodrigues de Lima}
\author[b]{Pedro Henrique Ribeiro da Silva Moraes}

\affiliation[a]{Universidade do Estado de Santa Catarina (UDESC), Joinville, SC, Brazil}
\affiliation[b]{N\'ucleo de Astrof\'isica Te\'orica, Universidade Cidade de S\~ao Paulo, Rua Galv\~ao Bueno 868, 01506-000, S\~ao Paulo, SP, Brazil}

\emailAdd{samuel.isidoro@example.com}
\emailAdd{rafael.lima@example.com}
\emailAdd{moraes.phrs@gmail.com}

\abstract{

We use Bayesian inference on multi-messenger observational data in order to constrain the parameter space of compact stars in a phenomenological braneworld model inspired by the effective Sahni--Shtanov scenario. We model stellar structure via modified Tolman--Oppenheimer--Volkoff equations incorporating local quadratic brane corrections and a phenomenological closure for the nonlocal Weyl sector, parametrized by the brane tension $\lambda$, the Weyl coupling $\alpha_{\mathcal{U}}$ and the Weyl equation-of-state parameter $w_{\mathcal{U}}$. For the SLy hadronic equation of state, we employ an affine-invariant ensemble Markov Chain Monte Carlo sampler on three complementary datasets: mass--radius posterior samples inferred from GW170817 tidal-deformability analyses (LIGO/Virgo) and the mass--radius posteriors of PSR~J0740$+$6620 and PSR~J1231$-$1411 from NICER X-ray timing. We adopt a prior centered at large brane tension and at the General Relativity values of the Weyl parameters. The posterior yields $\log_{10}(\lambda/\mathrm{km}^{-2}) = 3.98^{+1.52}_{-1.44}$ (68\%), placing the preferred configurations in a large-tension regime where local high-energy corrections are subdominant. At the posterior median, the shift of the stellar sequence is driven mainly by the Weyl sector rather than by the local quadratic term. The Weyl coupling is constrained to $\alpha_{\mathcal{U}} = -0.15^{+0.30}_{-0.08}$ (68\%), with a preference for slightly negative values that reduce the effective gravitational density and expand stellar radii. The Weyl equation-of-state parameter $w_{\mathcal{U}}$ is only weakly constrained by the data, with a broad posterior spanning $[-1.25,\,1.02]$ (95\%). The inferred derived {\bf stellar} observables are $M_{\max} = 2.30^{+0.14}_{-0.08}\,M_\odot$ and $R_{1.4} = 13.31^{+0.54}_{-0.57}\,\mathrm{km}$ (68\%), both larger than the General Relativity predictions for the SLy equation of state ($M_{\max}^{\rm GR}\simeq2.05\,M_\odot$, $R_{1.4}^{\rm GR}\simeq11.76\,\mathrm{km}$). The 95\% posterior interval reaches the GW190814 secondary-mass range, although the median and best-fit values remain below it. These results establish that a large-tension braneworld with a mildly negative Weyl coupling produces compact-star sequences consistent with current NICER and gravitational-wave constraints without requiring extreme departures from General Relativity.
}


\arxivnumber{}

\begin{document}

\maketitle
\flushbottom

\section{Introduction}

Compact stars provide one of the most powerful probes of gravity in the strong-field regime. Recent observations of neutron stars with masses near or above $2M_\odot$ \cite{Demorest2010,Antoniadis2013,Cromartie2020,Romani2025} and mass--radius measurements from NICER X-ray timing \cite{Brandes2025} impose stringent constraints on both the nuclear equation of state (EoS) and possible modifications of General Relativity (GR). The secondary component of GW190814 is especially relevant in this context: if interpreted as a neutron star, its inferred mass lies in the interval $2.50$--$2.67\,M_\odot$ at 90\% credibility \cite{Abbott2020GW190814}. Although its nature remains uncertain and a low-mass black hole interpretation is possible, GW190814 provides a useful benchmark for assessing how close a model comes to supporting very massive compact stars.

The interpretation of these observations is limited by the uncertain microphysics of matter at supranuclear densities. Above nuclear saturation density, the composition of cold beta-equilibrated matter may be affected by many-body nuclear forces, the density dependence of the symmetry energy, hyperons, meson condensates, or transitions to deconfined quark matter. These possibilities generate a wide range of pressure-density relations and therefore different predictions for radii, tidal deformabilities and maximum masses \cite{LattimerPrakash2007,Oertel2017,Brandes2025}. Multi-messenger data from massive radio pulsars, NICER X-ray timing and gravitational waves have narrowed the allowed EoS space, but they have not removed the degeneracy between dense-matter physics and strong-field gravity.

This degeneracy motivates controlled tests in which the EoS is fixed while the gravitational sector is varied. GR has passed all available weak-field and binary-pulsar tests with high precision, but the interior of a neutron star probes non-vacuum curvature and compactness regimes that are inaccessible in the Solar System. Modified-gravity scenarios can therefore remain perturbative in weak fields while producing observable shifts in stellar structure, maximum masses, or tidal response \cite{Berti2015}. Compact stars are consequently a natural laboratory for testing whether deviations from GR can mimic, enhance, or compete with changes in the high-density EoS.

Modifications of GR have been widely proposed in the last decades, and are mainly motivated by the recent accelerated expansion the Universe is going through \cite{riess/1998,rubin/2016,weinberg/2013}. While GR carries the infamous {\it cosmological constant problem} \cite{weinberg/1989}, modified theories of gravity can evade or alleviate such a high discrepancy between theoretical and observable values of the quantum vacuum energy density.

Among the most studied modified-gravity scenarios are braneworld models, originally proposed by Randall and Sundrum \cite{Randall1999a,Randall1999b}, in which ordinary matter is confined to a four-dimensional hypersurface while gravity propagates in a higher-dimensional bulk. Effective gravitational equations on the brane were derived by Shiromizu, Maeda, and Sasaki \cite{Shiromizu2000}, leading naturally to local quadratic corrections and nonlocal Weyl contributions.

Compact stars in braneworld gravity have been investigated in several contexts \cite{Germani2001,Ovalle2009,Ovalle2013,Castro2014,GarciaAspeitia2015,Murshid2025}. In particular, modifications of the Tolman--Oppenheimer--Volkoff (TOV) equations may significantly alter stellar equilibrium properties, including the maximum mass and compactness.

In this work we study compact stars in a phenomenological effective braneworld framework inspired by the Sahni--Shtanov scenario \cite{Sahni2003,Mishra2025}. The model should be understood as a four-dimensional closure of the Shiromizu--Maeda--Sasaki effective equations rather than as a complete five-dimensional Sahni--Shtanov bulk solution. We derive modified TOV equations including both local quadratic corrections and phenomenological Weyl terms, and numerically solve the resulting system for the SLy hadronic equation of state \cite{Douchin2001}. Keeping the EoS fixed allows us to isolate the effect of the braneworld sector on the mass--radius relation and on the inferred maximum mass.

We perform a multi-messenger Bayesian inference of this effective braneworld parametrization using neutron-star observations. The posterior identifies a large-tension regime in which the local high-energy correction is subdominant and a slightly negative Weyl coupling $\alpha_{\mathcal{U}}<0$ reduces the effective gravitational density, shifting mass--radius sequences to larger radii and maximum masses relative to GR. This combination is consistent with current multi-messenger constraints and establishes the Weyl sector as the dominant phenomenological lever in this effective framework.

\section{Effective braneworld equations}

The effective Einstein equations on the brane can be written as \cite{Shiromizu2000}

\begin{equation}
G_{\mu\nu}
=
8\pi T_{\mu\nu}
+
\kappa_5^4 S_{\mu\nu}
-
E_{\mu\nu},
\label{eq:braneeinstein}
\end{equation}
where \(S_{\mu\nu}\) contains local quadratic corrections in the brane energy-momentum tensor, \(E_{\mu\nu}\) is the projected bulk Weyl tensor, and \(\kappa_5\) is the five-dimensional gravitational coupling constant, related to the five-dimensional Newton constant by \(\kappa_5^2=8\pi G_5\).

For a static spherically symmetric fluid, we consider effective density and pressure terms of the form

\begin{equation}
\rho_{\mathrm{eff}}
=
\rho
+
\frac{\rho^2}{2\lambda}
+
\rho_{\mathcal U},
\label{eq:rhoeff}
\end{equation}

\begin{equation}
p_{\mathrm{eff}}
=
p
+
\frac{p\rho}{\lambda}
+
\frac{\rho^2}{2\lambda}
+
p_{\mathcal U},
\label{eq:peff}
\end{equation}

where $\lambda$ is the brane tension parameter.

For the Weyl sector we adopt the phenomenological parametrization

\begin{equation}
\rho_{\mathcal U}
=
\alpha_{\mathcal U}\rho,
\label{eq:rhoU}
\end{equation}

\begin{equation}
p_{\mathcal U}
=
w_{\mathcal U}\rho_{\mathcal U}.
\label{eq:pU}
\end{equation}

The modified TOV equations are

\begin{equation}
\frac{dm}{dr}
=
4\pi r^2 \rho_{\mathrm{eff}},
\label{eq:mass}
\end{equation}

and

\begin{equation}
\frac{dp}{dr}
=
-(\rho+p)
\frac{
m+4\pi r^3 p_{\mathrm{eff}}
}
{
r(r-2m)
}.
\label{eq:tov}
\end{equation}

In the limit

\begin{equation}
\lambda\rightarrow\infty,
\qquad
\alpha_{\mathcal U}\rightarrow0,
\label{eq:grlimit}
\end{equation}

the standard GR TOV equations are recovered.

\section{Equation of state}

We employ the realistic SLy equation of state, developed by Douchin and Haensel \cite{Douchin2001}, which provides a unified description of neutron-star matter from the outer crust to the dense core. The EoS tabulated data are implemented numerically using monotonic PCHIP spline interpolation.

\section{Bayesian inference setup}

\subsection{TOV integration}

We integrate the modified TOV equations~\eqref{eq:mass}--\eqref{eq:tov} using an adaptive fourth-order Runge--Kutta method. Stellar sequences are generated on a grid of $N_\rho = 120$ central densities logarithmically spaced in
\begin{equation}
10^{14}
\lesssim
\rho_c
\lesssim
10^{16}\ \mathrm{g/cm^3}.
\label{eq:rhorange}
\end{equation}
For each parameter set $\theta = (\lambda,\,\alpha_{\mathcal{U}},\,w_{\mathcal{U}})$ the stable branch of the mass--radius relation is extracted as the sequence of models up to the maximum-mass configuration.

\subsection{MCMC sampler}

We infer the braneworld parameters using an affine-invariant ensemble Markov Chain Monte Carlo (MCMC) sampler~\cite{ForemanMackey2013}. We run $N_e = 4$ independent parallel ensembles, each with $N_w = 32$ walkers and $N_s = 3000$ steps, using the stretch proposal move. The first $25\%$ of steps ($N_{\rm burn} = 750$) are discarded as burn-in, yielding $N_{\rm eff} = N_e \times N_w \times (N_s - N_{\rm burn}) = 288\,000$ posterior samples. The global acceptance rate is $49\%$. Convergence is assessed via the Gelman--Rubin statistic; $\log_{10}\lambda$ satisfies $\hat{R} \approx 1.002$, while $\alpha_{\mathcal{U}}$ and $w_{\mathcal{U}}$ reach $\hat{R} \approx 1.02$, consistent with their posterior degeneracy.

\subsection{Observational datasets and likelihood}

We constrain the model using three complementary multi-messenger datasets:

\begin{enumerate}

\item \textit{GW170817}: mass--radius posterior samples derived from LIGO/Virgo tidal deformability measurements of the binary neutron-star merger~\cite{Abbott2018}. The likelihood compares the stable $R(M)$ branch against the observed $(m_1,\,R_1)$ and $(m_2,\,R_2)$ pairs using a Gaussian KDE in radius with bandwidth $h = 1.2\,\mathrm{km}$ and weight $w_{\rm GW} = 0.3$.

\item \textit{NICER PSR J0740$+$6620}: mass--radius posterior samples from the updated NICER and XMM-Newton X-ray timing analysis~\cite{Dittmann2024}, evaluated with bandwidth $h = 0.85\,\mathrm{km}$ and weight $w_{\rm J0740} = 1.5$.

\item \textit{NICER PSR J1231$-$1411}: mass--radius posterior samples from NICER X-ray timing~\cite{Salmi2024J1231}, evaluated with the same bandwidth $h = 0.85\,\mathrm{km}$ and weight $w_{\rm J1231} = 1.25$. We use the PDTU\_H\_R1014\footnote{The label PDTU\_H designates the PDT-U waveform model, Pulse Decomposition Technique with unconstrained hot-spot geometry, combined with a hydrogen atmosphere, following the naming convention of Ref.~\cite{Salmi2024J1231}. The suffix R1014 denotes the uniform radius prior restricted to $[10,\,14]\,\mathrm{km}$.} posterior, obtained with an uninformative radius prior restricted to $10$--$14\,\mathrm{km}$.

\end{enumerate}

For each dataset the log-likelihood contribution is computed as
\begin{equation}
\ln \mathcal{L}_k
=
\ln \!\left[ \sum_i \exp\!\left(-\frac{[R_{\rm model}(M_i) - R_i]^2}{2h^2}\right) \right]
+
\kappa \ln f_{\rm cov},
\label{eq:loglike}
\end{equation}
where the sum runs over the observational samples, $R_{\rm model}(M_i)$ is the model radius interpolated at each observed mass, $f_{\rm cov}$ is the fraction of the observed mass--radius support covered by the stable model branch, and $\kappa = 8$ is the support-coverage penalty weight. The total log-posterior is
\begin{equation}
\ln \mathcal{P}
=
\ln \pi
+
\sum_k w_k \ln \mathcal{L}_k
+
\ln \mathcal{P}_{\rm constr},
\label{eq:logposterior}
\end{equation}
where $\ln \pi$ is the log-prior and $\ln \mathcal{P}_{\rm constr}$ encodes soft physical constraints: a Gaussian penalty for $M_{\max} < 2.12\,M_\odot$ (width $\sigma = 0.06\,M_\odot$) and a penalty for $C_{\max} > 0.34$ (width $\sigma = 0.05$).

\subsection{Priors}

We adopt the large-tension prior set, consisting of a Gaussian prior centered at large brane tension for $\log_{10}\lambda$ and Gaussian priors centered on the General Relativity values of the Weyl parameters, multiplied by uniform hard bounds:
\begin{align}
\log_{10}(\lambda/\mathrm{km}^{-2}) &\sim \mathcal{N}(4,\;1.5^2) \times \mathcal{U}(-1,\;9), \label{eq:prior_lambda} \\
\alpha_{\mathcal{U}} &\sim \mathcal{N}(0,\;0.25^2) \times \mathcal{U}(-1.5,\;1.5), \label{eq:prior_alpha} \\
w_{\mathcal{U}} &\sim \mathcal{N}(0,\;0.5^2) \times \mathcal{U}(-2,\;2). \label{eq:prior_w}
\end{align}
The Gaussian component on $\log_{10}\lambda$ regularizes the exploration toward the large-tension regime, while the components on $\alpha_{\mathcal{U}}$ and $w_{\mathcal{U}}$ enforce a mild preference for the GR Weyl sector values. Consequently, posterior structure in $\lambda$ should be interpreted together with prior sensitivity, whereas robust shifts in the mass--radius sequence are expected primarily through the Weyl parameters when the local quadratic term is small.

\section{Results}

\subsection{Posterior parameter constraints}

\begin{figure}[htbp]
\centering
\includegraphics[width=0.95\textwidth]{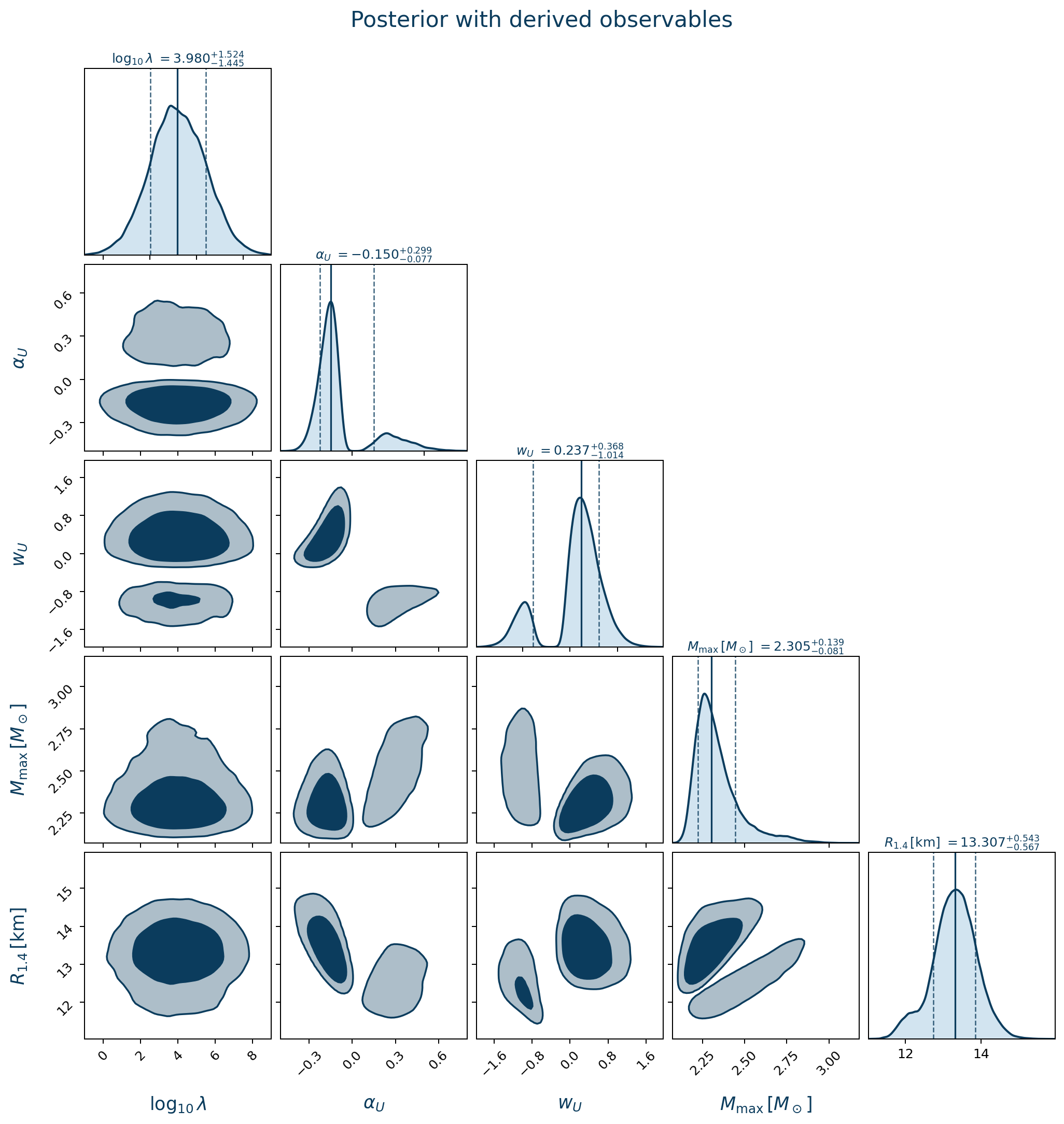}
\caption{Posterior distributions for the braneworld parameters $(\log_{10}\lambda,\,\alpha_{\mathcal{U}},\,w_{\mathcal{U}})$ and derived observables $(M_{\max},\,R_{1.4})$, obtained from the MCMC fit to LIGO/Virgo GW170817 and NICER PSR~J0740$+$6620 and PSR~J1231$-$1411 data after discarding the first 25\% of steps from each ensemble. Contours enclose 68\% and 95\% of the posterior probability. Diagonal panels show marginalized one-dimensional posteriors; vertical dashed lines mark the 16th, 50th, and 84th percentiles. The smaller positive-$\alpha_{\mathcal U}$, negative-$w_{\mathcal U}$ structure corresponds to a subdominant Weyl-compensation branch discussed in the text.}
\label{fig:corner}
\end{figure}

Figure~\ref{fig:corner} shows the joint and marginalized posterior distributions for the braneworld parameters $(\log_{10}\lambda,\,\alpha_{\mathcal{U}},\,w_{\mathcal{U}})$ together with the derived observables $(M_{\max},\,R_{1.4})$, computed from the post-burn-in chains defined above. The posterior results are summarized in Table~\ref{tab:posterior}.

\begin{table}[ht]
\centering
\caption{Posterior summary for the braneworld parameters and derived observables. Uncertainties are 68\% and 95\% credible intervals (CI). The GR reference values for the SLy EoS are listed for comparison.}
\label{tab:posterior}
\begin{tabular}{lcccc}
Quantity & Median & 68\% CI & 95\% CI & GR (SLy) \\
\hline
$\log_{10}(\lambda/\mathrm{km}^{-2})$ & $3.98$ & $[2.54,\;5.50]$ & $[1.10,\;6.93]$ & $\infty$ \\
$\alpha_{\mathcal{U}}$                & $-0.15$ & $[-0.23,\;0.15]$ & $[-0.31,\;0.44]$ & $0$ \\
$w_{\mathcal{U}}$                     & $0.24$  & $[-0.78,\;0.61]$ & $[-1.25,\;1.02]$ & $0$ \\
$M_{\max}$ [$M_\odot$]               & $2.305$ & $[2.223,\;2.444]$ & $[2.169,\;2.721]$ & $2.05$ \\
$R_{1.4}$ [km]                        & $13.31$ & $[12.74,\;13.85]$ & $[11.97,\;14.42]$ & $11.76$ \\
\end{tabular}
\end{table}

The posterior median for the brane tension is $\log_{10}(\lambda/\mathrm{km}^{-2}) = 3.98^{+1.52}_{-1.44}$ (68\%), corresponding to $\lambda \approx 10^{4}\,\mathrm{km}^{-2}$. This places the preferred configurations in the large-tension regime favored by the adopted prior. At the median value, typical stellar central densities correspond to $\rho/\lambda \sim 10^{-7}$ in the geometric units used in the integration, so the local quadratic correction $\rho^2/(2\lambda)$ is strongly suppressed. The observational leverage in this setup therefore comes primarily from the Weyl sector, especially $\alpha_{\mathcal U}$ and the product $w_{\mathcal U}\alpha_{\mathcal U}$, while $\lambda$ controls a subdominant local high-energy contribution.

The Weyl coupling is constrained to $\alpha_{\mathcal{U}} = -0.15^{+0.30}_{-0.08}$ (68\%), with 83\% of the posterior mass at $\alpha_{\mathcal{U}}<0$. A negative $\alpha_{\mathcal{U}}$ reduces the effective gravitational density via $\rho_{\rm eff} = \rho(1+\alpha_{\mathcal{U}})$, expanding the stellar radius relative to GR. The Weyl equation-of-state parameter $w_{\mathcal{U}}$ is only weakly constrained by the data: its posterior spans most of the prior range, with a 95\% CI of $[-1.25,\;1.02]$, indicating near-complete degeneracy in this direction.

The corner plot also shows a subdominant branch with positive $\alpha_{\mathcal U}$ and negative $w_{\mathcal U}$, containing about 17\% of the post-burn-in posterior samples. This branch is not an independent high-probability solution but a compensation allowed by the phenomenological Weyl closure. Positive $\alpha_{\mathcal U}$ increases the effective density in Eq.~\eqref{eq:mass}, while a negative value of the product $w_{\mathcal U}\alpha_{\mathcal U}$ reduces the effective radial pressure contribution entering Eq.~\eqref{eq:tov}. The resulting configurations typically have smaller $R_{1.4}$, larger $M_{\max}$, and compactness closer to the imposed upper bound than the dominant negative-$\alpha_{\mathcal U}$ branch. We therefore interpret this feature as a lower-probability Weyl-compensation branch rather than as a robust separate prediction of the model.

The best post-burn-in posterior sample yields $\lambda = 10408\,\mathrm{km}^{-2}$ ($\log_{10}\lambda = 4.02$), $\alpha_{\mathcal{U}} = -0.169$, $w_{\mathcal{U}} = 0.252$, with $M_{\max} = 2.277\,M_\odot$ and $R_{1.4} = 13.29\,\mathrm{km}$.

It is useful to compare these values with the secondary component of GW190814. If that object is interpreted as a neutron star, its mass interval, $2.50$--$2.67\,M_\odot$ at 90\% credibility \cite{Abbott2020GW190814}, remains above both the posterior median and the best-fit maximum mass obtained here. However, the 95\% credible interval, $M_{\max}=[2.169,\;2.721]\,M_\odot$, overlaps the GW190814 interval through the upper posterior tail. The present model should therefore be viewed as allowing GW190814-like masses only in the high-mass tail, rather than as a robust median explanation of GW190814.

\subsection{Mass--radius posterior band}

\begin{figure}[htbp]
\centering
\includegraphics[width=0.9\textwidth]{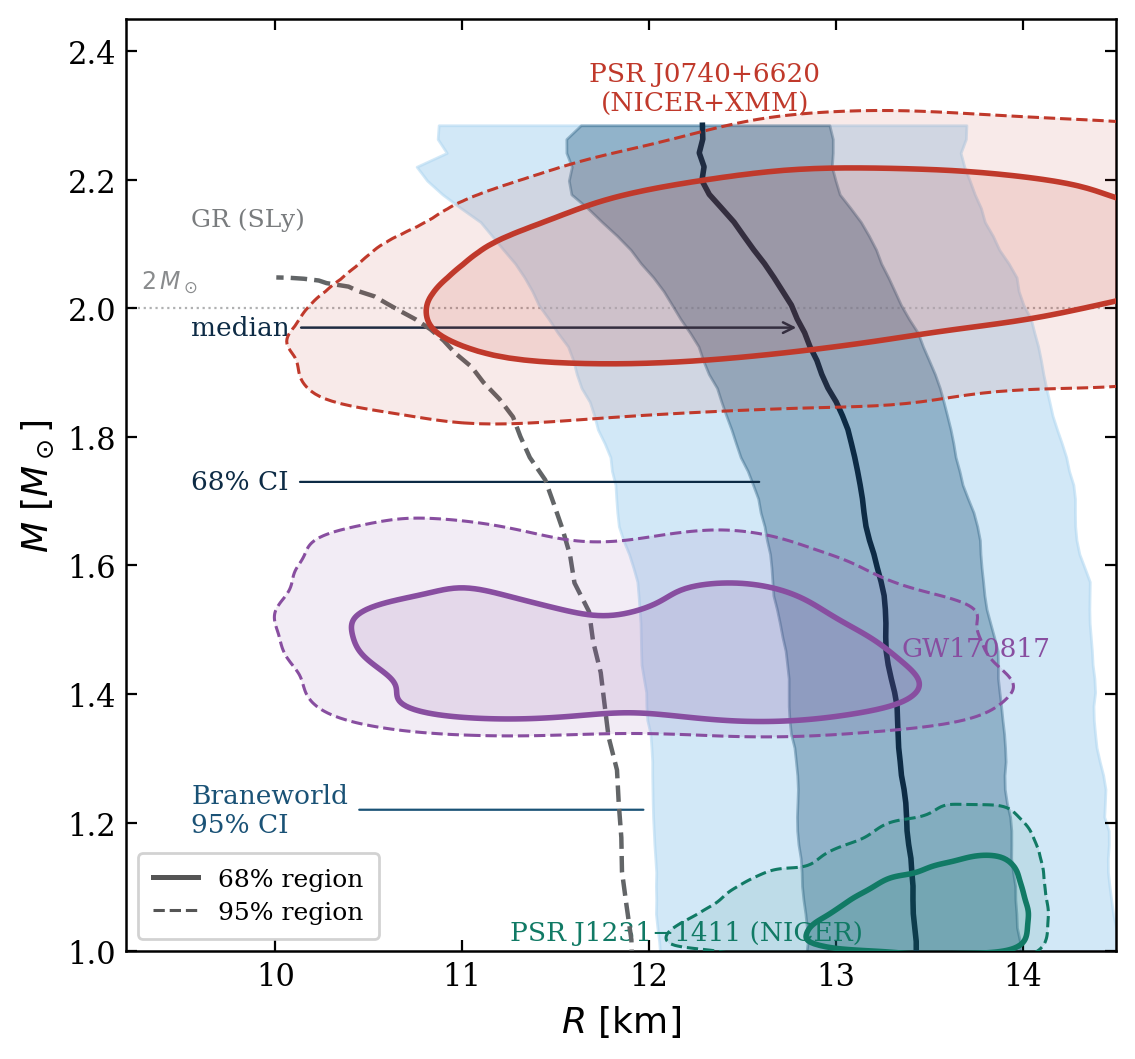}
\caption{Posterior predictive mass--radius band for the SLy equation of state in the effective Sahni--Shtanov braneworld model. Shaded regions correspond to 68\% (dark blue) and 95\% (light blue) credible bands. The dashed curve shows the standard GR result for comparison. The posterior is constrained by LIGO/Virgo GW170817 tidal deformability and NICER mass--radius measurements of PSR~J0740$+$6620 and PSR~J1231$-$1411.}
\label{fig:mr}
\end{figure}

Figure~\ref{fig:mr} shows a posterior mass--radius band, computed from a random representative subset of 1000 post-burn-in samples. The band is intended as a visual posterior summary rather than a high-precision posterior-predictive emulator. The GR prediction for the SLy EoS ($M_{\max}^{\rm GR} \simeq 2.05\,M_\odot$) is shown for comparison.

The median posterior radius at $1.4\,M_\odot$ is $R_{1.4} = 13.31\,\mathrm{km}$, larger than the GR SLy value of $11.76\,\mathrm{km}$. The posterior band overlaps with the NICER and LIGO/Virgo mass--radius constraints used here, indicating internal consistency of this phenomenological scenario with the adopted observational data. The rightward shift relative to GR is driven by the slightly negative $\alpha_{\mathcal{U}}$, which reduces the effective mass enclosed within each radius via the modified mass equation~\eqref{eq:mass}.

\subsection{Physical interpretation}

The posterior-preferred regime combines two main effects:

\begin{enumerate}

\item \textit{Large effective brane tension}: $\lambda \sim 10^4\,\mathrm{km}^{-2}$ suppresses the local quadratic correction $\rho^2/(2\lambda)$, keeping the local high-energy term perturbative and close to the GR limit. Because the posterior in $\lambda$ remains close to the imposed prior, this should be read mainly as a large-tension compatible regime rather than as a sharp observational determination of the brane tension.

\item \textit{Slightly negative Weyl coupling}: $\alpha_{\mathcal{U}} \approx -0.15$ reduces the effective gravitational density via $\rho_{\rm eff} = \rho(1+\alpha_{\mathcal{U}}) < \rho$ and the enclosed mass via Eq.~\eqref{eq:mass}, expanding stellar radii and raising the maximum mass relative to GR.

\end{enumerate}

The Weyl pressure parameter $w_{\mathcal{U}}$ is found to be nearly unconstrained by the current data, with a posterior that spans most of the prior range. This degeneracy reflects the fact that $w_{\mathcal{U}}$ enters the TOV equation~\eqref{eq:tov} only through the combination $w_{\mathcal{U}}\alpha_{\mathcal{U}}\rho$, which is suppressed when $|\alpha_{\mathcal{U}}|$ is small. Future observations with tighter NICER statistics or next-generation gravitational-wave detectors may break this degeneracy.

The same degeneracy explains the secondary positive-$\alpha_{\mathcal U}$ branch visible in the corner plot. In that region the Weyl density term and Weyl pressure term pull the stellar sequence in partially opposite directions, allowing an acceptable but less favored fit. Its lower posterior weight and proximity to the compactness penalty indicate that the data and soft physical constraints prefer the simpler screening interpretation associated with $\alpha_{\mathcal U}<0$.

The net result is that the braneworld posterior predicts $M_{\max} \approx 2.30\,M_\odot$ and $R_{1.4} \approx 13.31\,\mathrm{km}$, both larger than the GR SLy predictions, and consistent with the observational constraints from NICER and LIGO/Virgo under the empirical mass--radius likelihood adopted here. This rightward shift of the mass--radius sequence with respect to GR is the expected signature of a Weyl contribution that partially screens the gravitational interaction inside the star.

\section{Comparison with previous braneworld compact-star studies}
\label{sec:discussion_literature}

One of the most relevant outcomes of the present analysis is that the increase of the maximum mass of compact stars is not a generic prediction of effective braneworld gravity. Instead, our results indicate that the behavior of the maximum mass depends critically on the implementation of the effective braneworld corrections, particularly on the treatment of the nonlocal Weyl sector.

A large part of the braneworld compact-star literature has emphasized the possibility that extra-dimensional effects may support heavier compact stars than those predicted by GR. This expectation appears naturally in scenarios where bulk effects weaken the effective four-dimensional gravitational interaction or induce repulsive contributions through the projected Weyl tensor \cite{Germani2001,Ovalle2009,Ovalle2013,Castro2014,GarciaAspeitia2015,Murshid2025}. In several analyses, modifications of the exterior geometry or anisotropic Weyl stresses lead to equilibrium configurations with masses exceeding the GR limit.

However, the present work shows that this behavior is highly model dependent. In our effective Sahni--Shtanov-inspired framework, the dominant contribution arises from the local quadratic correction

\begin{equation}
\rho_{\mathrm{loc}}
=
\rho+\frac{\rho^2}{2\lambda},
\label{eq:rholocaldiscussion}
\end{equation}
\noindent
which systematically increases the effective gravitational density inside the star. Considered without a compensating Weyl contribution, this term tends to destabilize configurations at lower masses and smaller radii compared to GR.

Even after including phenomenological Weyl contributions through {\bf \eqref{eq:rhoU} and \eqref{eq:pU},} the local high-energy correction alone would generally reduce the maximum mass relative to the GR prediction for the realistic SLy equation of state. In contrast, a negative $\alpha_{\mathcal{U}}$ contributes negative effective Weyl density and can compensate this enhanced attraction. The posterior-preferred value $\alpha_{\mathcal{U}}\simeq -0.15$ is precisely in this screening regime and shifts the mass--radius sequence above the GR+SLy reference.

This result is particularly important because it demonstrates explicitly that effective Weyl parametrizations do not automatically generate heavier compact stars. In the present model, the sign and magnitude of the Weyl contribution determine whether local quadratic corrections remain dominant or are partially screened.

The situation is further complicated by the fact that the Weyl sector is not uniquely determined by the four-dimensional effective equations. As emphasized by Shiromizu, Maeda, and Sasaki \cite{Shiromizu2000}, the projected Weyl tensor carries information about the bulk geometry that cannot be reconstructed solely from brane quantities. Consequently, any phenomenological closure of the Weyl sector necessarily introduces an additional level of model dependence.

This observation may help explain why different braneworld studies frequently obtain qualitatively different compact-star phenomenologies. Depending on the assumed bulk geometry, exterior matching conditions, anisotropic stresses or Weyl closure relations, the resulting stellar configurations may display either enhanced or suppressed maximum masses.

In this sense, our results suggest that claims of naturally larger compact-star masses in braneworld gravity should be interpreted with caution. The emergence of heavier stars does not appear to be a robust consequence of the effective braneworld framework itself, but rather a feature associated with specific assumptions about the bulk-induced corrections.

The present analysis therefore provides a complementary perspective to previous studies. Instead of assuming that extra-dimensional corrections necessarily increase the maximum mass, we show within the SLy hadronic setup considered here that local braneworld corrections and Weyl screening compete directly in the stellar equilibrium equations.

These results reinforce the idea that fully self-consistent bulk dynamics may be essential for obtaining realistic braneworld compact-star configurations compatible with current astrophysical constraints from massive pulsars, NICER measurements, and gravitational-wave observations.

The present Bayesian analysis identifies precisely such a viable window: a large-tension regime combined with slightly negative $\alpha_{\mathcal{U}}$ produces stellar sequences consistent with current multi-messenger data without requiring extreme parameter values. This regime complements the earlier fixed-parameter studies and provides a data-driven anchor for future theoretical investigations.

\section{Discussion}

Our Bayesian analysis shows that multi-messenger data from GW170817 and NICER are compatible with a large-tension effective braneworld regime and a slightly negative Weyl coupling ($\alpha_{\mathcal{U}} \approx -0.15$). In this regime the local quadratic term controlled by $\lambda$ is subdominant, while the phenomenologically relevant shift of the stellar sequence is driven mainly by the Weyl coupling. This combination yields $M_{\max} \approx 2.30\,M_\odot$ and $R_{1.4} \approx 13.31\,\mathrm{km}$, both exceeding the GR+SLy predictions and consistent with current multi-messenger constraints under the adopted empirical likelihood.

The slightly negative $\alpha_{\mathcal{U}}$ produces a mild gravitational screening: by reducing the effective density entering both the mass equation and the TOV pressure gradient, it shifts the mass--radius sequence to larger radii and maximum masses. This is the physically expected direction of braneworld corrections when the Weyl sector provides a net negative contribution to the effective source.

The inferred maximum mass is also compatible within $1\sigma$ with the refined mass of PSR~J0952$-$0607, $M=2.35\pm0.11\,M_\odot$ \cite{Romani2025}. This comparison provides an additional high-mass pulsar benchmark that is less extreme than GW190814 but more constraining than the canonical $2\,M_\odot$ threshold.

The GW190814 comparison sharpens the interpretation of this mass increase. Relative to the GR+SLy reference value, the posterior median raises $M_{\max}$ by about $0.25\,M_\odot$, and the 95\% posterior interval reaches the GW190814 secondary-mass range. Nevertheless, the 68\% credible interval remains below that range. Any claim that the present SLy-based braneworld model explains the GW190814 secondary as a neutron star would therefore require stronger evidence, a broader EoS study, or additional physics beyond the effective parametrization adopted here.

From Fig.~\ref{fig:mr}, it should be remarked that, within the adopted fixed-SLy comparison, the posterior-preferred braneworld sequence shows better visual consistency with the combined GW170817, PSR~J0740$+$6620, and PSR~J1231$-$1411 mass--radius regions than the GR+SLy reference curve. It highlights that the effective Weyl screening can shift a soft SLy baseline into the region preferred by the adopted multi-messenger constraints.

The Weyl equation-of-state parameter $w_{\mathcal{U}}$ is poorly constrained by the current data, with a posterior largely driven by the prior. This degeneracy arises because $w_{\mathcal{U}}$ only enters through the product $w_{\mathcal{U}}\alpha_{\mathcal{U}}$, which is suppressed for small $|\alpha_{\mathcal{U}}|$. The Gelman--Rubin statistic for $\alpha_{\mathcal{U}}$ and $w_{\mathcal{U}}$ reaches $\hat{R} \approx 1.02$, slightly above the ideal convergence threshold of 1.01, consistent with this degeneracy rather than a sampling failure.

Future work should include:

\begin{enumerate}

\item self-consistent bulk solutions connecting the brane parametrization to a fully five-dimensional geometry;

\item anisotropic Weyl stress tensors beyond the isotropic phenomenological closure used here;

\item radial stability analyses of the braneworld stellar models;

\item tidal deformability calculations for comparison with future gravitational-wave events.

\end{enumerate}

\section{Conclusions}

We performed a multi-messenger Bayesian MCMC inference of a phenomenological Sahni--Shtanov-inspired braneworld parametrization using gravitational-wave (GW170817) and X-ray timing (NICER PSR~J0740$+$6620 and PSR~J1231$-$1411) data.

Our main conclusions are:

\begin{enumerate}

\item The posterior favors $\log_{10}(\lambda/\mathrm{km}^{-2}) = 3.98^{+1.52}_{-1.44}$ (68\%), corresponding to a large-tension regime in which local high-energy corrections are subdominant.

\item The Weyl coupling is constrained to $\alpha_{\mathcal{U}} = -0.15^{+0.30}_{-0.08}$ (68\%), with 83\% of the posterior at $\alpha_{\mathcal{U}} < 0$, indicating a slight gravitational screening by the Weyl sector.

\item The Weyl equation-of-state parameter $w_{\mathcal{U}}$ is not meaningfully constrained by the current data; its posterior spans most of the prior range, reflecting a degeneracy with $\alpha_{\mathcal{U}}$ in the TOV equations.

\item The inferred stellar observables $M_{\max} = 2.30^{+0.14}_{-0.08}\,M_\odot$ and $R_{1.4} = 13.31^{+0.54}_{-0.57}\,\mathrm{km}$ (68\%) both exceed the GR+SLy predictions, consistent with the observed multi-messenger data.

\item The inferred maximum mass is consistent within $1\sigma$ with the refined high-mass pulsar measurement of PSR~J0952$-$0607, $M=2.35\pm0.11\,M_\odot$.

\item The mass--radius sequence of the posterior-preferred braneworld model shifts to the right of the GR curve, confirming that the effective Weyl sector can expand neutron stars beyond the General Relativity limit without violating current observational constraints.

\item The inferred maximum mass reaches the GW190814 secondary-mass interval only through the upper 95\% posterior tail. Thus, within the fixed SLy setup, the model alleviates but does not turn a possible $2.6\,M_\odot$ neutron-star interpretation of GW190814 into a median prediction.

\end{enumerate}

These results establish that a large-tension braneworld with a mildly negative Weyl coupling produces compact-star sequences consistent with current NICER and gravitational-wave data. Within this effective closure, the Weyl sector rather than the local brane correction is the dominant lever through which effective braneworld gravity departs from GR in the compact-star mass--radius observable space.

\acknowledgments

RCRL acknowledge the support of the Universidade do Estado de Santa Catarina (UDESC) and the Fundação de Amparo à Pesquisa e Inovação do Estado de Santa Catarina (FAPESC). This work was partially supported by these institutions. PHRSM would like to thank CNPq (Conselho Nacional de Desenvolvimento Científico e Tecnológico) for partial financial support under grant No. 310366/2023-2. The authors would like to acknowledge the late Prof. Manuel Malheiro for valuable discussions during the early stages of this project. His interest in the braneworld scenario and his insightful suggestions helped motivate part of the investigation reported in this work.

\appendix

\section[Modified TOV equation in braneworld gravity]{Derivation of the Modified Tolman--Oppenheimer--Volkoff Equation in Braneworld Gravity}
\label{app:tov_modificada}

In this appendix we present the derivation of the modified Tolman--Oppenheimer--Volkoff (TOV) equation used throughout this work. We consider the effective gravitational equations on the brane in the form
\begin{equation}
G_{\mu\nu}
=
8\pi T^{\rm eff}_{\mu\nu},
\label{eq:einstein_effective_appendix}
\end{equation}
where the effective energy--momentum tensor is written as
\begin{equation}
T^{\rm eff}_{\mu\nu}
=
T_{\mu\nu}
+
\frac{1}{\lambda}S_{\mu\nu}
+
\mathcal{E}_{\mu\nu}.
\label{eq:effective_tensor_appendix}
\end{equation}
Here, $\lambda$ is the brane tension, $S_{\mu\nu}$ contains the local quadratic corrections in the matter energy--momentum tensor, and $\mathcal{E}_{\mu\nu}$ is the projection of the bulk Weyl tensor onto the brane.

We assume a static and spherically symmetric metric,
\begin{equation}
ds^2
=
-e^{\Phi(r)}dt^2
+
\left(1-\frac{2m(r)}{r}\right)^{-1}dr^2
+
r^2d\Omega^2,
\label{eq:metric_appendix}
\end{equation}
and model ordinary matter as a perfect fluid,
\begin{equation}
T^{\mu}_{\ \nu}
=
\mathrm{diag}(-\rho,p,p,p).
\label{eq:perfect_fluid_appendix}
\end{equation}

The local quadratic corrections modify the effective density and pressure according to
\begin{equation}
\rho_{\rm loc}
=
\rho
+
\frac{\rho^2}{2\lambda},
\label{eq:rho_local_appendix}
\end{equation}
and
\begin{equation}
p_{\rm loc}
=
p
+
\frac{p\rho}{\lambda}
+
\frac{\rho^2}{2\lambda}.
\label{eq:p_local_appendix}
\end{equation}

The Weyl contribution is parametrized by an effective density $\rho_{\mathcal U}$ and by radial and tangential effective pressures, $p_{\mathcal U}$ and $p_{t,\mathcal U}$. Therefore, we define
\begin{equation}
\rho_{\rm tot}
=
\rho_{\rm loc}
+
\rho_{\mathcal U},
\label{eq:rho_total_appendix}
\end{equation}
\begin{equation}
p_{r,{\rm tot}}
=
p_{\rm loc}
+
p_{\mathcal U},
\label{eq:pr_total_appendix}
\end{equation}
and
\begin{equation}
p_{t,{\rm tot}}
=
p_{\rm loc}
+
p_{t,\mathcal U}.
\label{eq:pt_total_appendix}
\end{equation}

The temporal component of Einstein's equations yields the modified mass equation,
\begin{equation}
\frac{dm}{dr}
=
4\pi r^2\rho_{\rm tot}.
\label{eq:mass_equation_appendix}
\end{equation}
Substituting the total effective density, we obtain
\begin{equation}
\frac{dm}{dr}
=
4\pi r^2
\left[
\rho
+
\frac{\rho^2}{2\lambda}
+
\rho_{\mathcal U}
\right].
\label{eq:mass_equation_explicit_appendix}
\end{equation}

The radial component of Einstein's equations can be written as
\begin{equation}
8\pi p_{r,{\rm tot}}
=
-\frac{1}{r^2}
+
\left(1-\frac{2m}{r}\right)
\left(
\frac{1}{r^2}
+
\frac{\Phi'}{r}
\right).
\label{eq:einstein_rr_appendix}
\end{equation}
Solving for $\Phi'(r)$ gives
\begin{equation}
\Phi'
=
\frac{
2\left[
m
+
4\pi r^3p_{r,{\rm tot}}
\right]
}
{r(r-2m)}.
\label{eq:phi_prime_appendix}
\end{equation}

Assuming separate covariant conservation of ordinary matter,
\begin{equation}
\nabla_\mu T^{\mu\nu}=0,
\label{eq:matter_conservation_appendix}
\end{equation}
we obtain the hydrostatic equilibrium equation
\begin{equation}
\frac{dp}{dr}
=
-\frac{1}{2}(\rho+p)\Phi'.
\label{eq:hydrostatic_appendix}
\end{equation}
Substituting Eq.~\eqref{eq:phi_prime_appendix}, we arrive at
\begin{equation}
\frac{dp}{dr}
=
-(\rho+p)
\frac{
m
+
4\pi r^3p_{r,{\rm tot}}
}
{r(r-2m)}.
\label{eq:tov_effective_appendix}
\end{equation}
Using the explicit form of $p_{r,{\rm tot}}$, the modified TOV equation becomes
\begin{equation}
\frac{dp}{dr}
=
-(\rho+p)
\frac{
m
+
4\pi r^3
\left[
p
+
\frac{p\rho}{\lambda}
+
\frac{\rho^2}{2\lambda}
+
p_{\mathcal U}
\right]
}
{r(r-2m)}.
\label{eq:tov_modified_appendix}
\end{equation}

Alternatively, if we impose conservation of the full effective source,
\begin{equation}
\nabla_\mu T_{\rm eff}^{\mu\nu}=0,
\label{eq:effective_conservation_appendix}
\end{equation}
we obtain the anisotropic TOV equation,
\begin{equation}
\frac{dp_{r,{\rm tot}}}{dr}
=
-
(\rho_{\rm tot}+p_{r,{\rm tot}})
\frac{
m
+
4\pi r^3p_{r,{\rm tot}}
}
{r(r-2m)}
+
\frac{2}{r}
\left(
p_{t,{\rm tot}}
-
p_{r,{\rm tot}}
\right).
\label{eq:anisotropic_tov_appendix}
\end{equation}
The last term explicitly represents the anisotropy induced by the Weyl sector.

To close the system phenomenologically, we adopt the parametrization
\begin{equation}
\rho_{\mathcal U}
=
\alpha_{\mathcal U}\rho,
\label{eq:rhoU_ansatz_appendix}
\end{equation}
\begin{equation}
p_{\mathcal U}
=
w_{\mathcal U}\rho_{\mathcal U},
\label{eq:pU_ansatz_appendix}
\end{equation}
and
\begin{equation}
p_{t,\mathcal U}
=
p_{\mathcal U}
+
\Delta_{\mathcal U}.
\label{eq:ptU_ansatz_appendix}
\end{equation}

Thus,
\begin{equation}
\rho_{\rm tot}
=
\rho
+
\frac{\rho^2}{2\lambda}
+
\alpha_{\mathcal U}\rho,
\label{eq:rho_total_ansatz_appendix}
\end{equation}
and
\begin{equation}
p_{r,{\rm tot}}
=
p
+
\frac{p\rho}{\lambda}
+
\frac{\rho^2}{2\lambda}
+
w_{\mathcal U}\alpha_{\mathcal U}\rho.
\label{eq:pr_total_ansatz_appendix}
\end{equation}

Therefore, the final mass equation becomes
\begin{equation}
\frac{dm}{dr}
=
4\pi r^2
\left[
\rho
+
\frac{\rho^2}{2\lambda}
+
\alpha_{\mathcal U}\rho
\right],
\label{eq:mass_final_appendix}
\end{equation}
while the modified TOV equation used in the model is
\begin{equation}
\frac{dp}{dr}
=
-(\rho+p)
\frac{
m
+
4\pi r^3
\left[
p
+
\frac{p\rho}{\lambda}
+
\frac{\rho^2}{2\lambda}
+
w_{\mathcal U}\alpha_{\mathcal U}\rho
\right]
}
{r(r-2m)}.
\label{eq:tov_final_appendix}
\end{equation}

In the limit
\begin{equation}
\lambda\rightarrow\infty,
\qquad
\alpha_{\mathcal U}\rightarrow0,
\qquad
w_{\mathcal U}\alpha_{\mathcal U}\rightarrow0,
\label{eq:gr_limit_appendix}
\end{equation}
the standard general relativistic TOV equation is recovered,
\begin{equation}
\frac{dp}{dr}
=
-(\rho+p)
\frac{
m
+
4\pi r^3p
}
{r(r-2m)}.
\label{eq:tov_gr_appendix}
\end{equation}

The terms proportional to $\rho^2/\lambda$ become important in the high-density regime and represent local high-energy corrections. The terms controlled by $\alpha_{\mathcal U}$ and $w_{\mathcal U}$ encode phenomenologically the nonlocal effects associated with the bulk geometry. Depending on the sign and magnitude of these parameters, the Weyl sector may either strengthen or weaken the effective gravitational interaction inside compact stars.

\bibliographystyle{JHEP}
\bibliography{references}

\end{document}